\documentclass[aps,pra,reprint,superscriptaddress,showpacs,letterpaper,longbibliography]{revtex4-1}
\usepackage[english]{babel}
\usepackage{ucs}
\usepackage[utf8x]{inputenc}
\usepackage{amsmath}
\usepackage{amsfonts}
\usepackage{amssymb}
\usepackage{graphicx}
\usepackage{bm}
\usepackage{bbm}
\usepackage{empheq}
\usepackage{color}
\makeatletter
\@ifundefined{textcolor}{}
{%
 \definecolor{BLACK}{gray}{0}
 \definecolor{WHITE}{gray}{1}
 \definecolor{RED}{rgb}{1,0,0}
 \definecolor{GREEN}{rgb}{0,1,0}
 \definecolor{BLUE}{rgb}{0,0,1}
 \definecolor{CYAN}{cmyk}{1,0,0,0}
 \definecolor{MAGENTA}{cmyk}{0,1,0,0}
 \definecolor{YELLOW}{cmyk}{0,0,1,0}
}

\definecolor{new}{rgb}{.8,.05,.08}

\begin{document}

\title{Injection Locking of a Semiconductor Double Quantum Dot Micromaser}

\author{Y.-Y. Liu}
\author{J. Stehlik}
\affiliation{Department of Physics, Princeton University, Princeton, New Jersey 08544, USA}
\author{M.~J.~Gullans}
\affiliation{Joint Quantum Institute, National Institute of Standards and Technology, Gaithersburg, Maryland 20899, USA}
\affiliation{Joint Center for Quantum Information and Computer Science, University of Maryland, College Park, Maryland 20742, USA}
\author{J.~M.~Taylor}
\affiliation{Joint Quantum Institute, National Institute of Standards and Technology, Gaithersburg, Maryland 20899, USA}
\affiliation{Joint Center for Quantum Information and Computer Science, University of Maryland, College Park, Maryland 20742, USA}
\author{J. R. Petta}
\affiliation{Department of Physics, Princeton University, Princeton, New Jersey 08544, USA}
\affiliation{Department of Physics, University of California, Santa Barbara, California 93106, USA}

\date{\today}

\begin{abstract}
 Emission linewidth is an important figure of merit for masers and lasers. We recently demonstrated a semiconductor double quantum dot (DQD) micromaser where photons are generated through single electron tunneling events. Charge noise directly couples to the DQD energy levels, resulting in a maser linewidth that is more than 100 times larger than the Schawlow-Townes prediction. Here we demonstrate a linewidth narrowing of more than a factor 10 by locking the DQD emission to a coherent tone that is injected to the input port of the cavity. We measure the injection locking range as a function of cavity input power and show that it is in agreement with the Adler equation. The position and amplitude of distortion sidebands that appear outside of the injection locking range are quantitatively examined. Our results show that this unconventional maser, which is impacted by strong charge noise and electron-phonon coupling, is well described by standard laser models.

\end{abstract}

%\pacs{73.21.La, 73.23.Hk, 84.40.lk}
% 73.21.La quantum dots
% 73.23.Hk Coulomb blockade; single-electron tunneling
% 84.40.lk Masers

\maketitle

\section{Introduction}

Masers and lasers have widespread applications in science and technology. Lasers are fundamentally different from conventional light sources (e.g.\ incandescent light bulbs and light emitting diodes) due to the fact that they emit coherent radiation of high intensity \cite{Siegman1986, Milonni1988}. These characteristics enable a wide range of applications, including industrial scale laser cutting and welding, light detection and ranging (LIDAR), and fundamental gravitational wave searches, such as the laser interferometer gravitational wave observatory (LIGO) \cite{Collis1976, Abramovici171992}. Laser interference patterns have a high contrast ratio and their application for cold atom trapping has opened up new frontiers in atomic, molecular, and optical physics (AMO) \cite{Prentiss1987, Aspect1988, Lett1988}.

In applications such as atomic clocks, interferometers, and optical lattices, it is important to have a narrow linewidth laser to achieve high spatial resolution and stability. Schawlow and Townes (ST) derived a minimum laser linewidth by considering the competition between spontaneous emission and stimulated emission \cite{Schawlow1958}.
In typical semiconductor lasers, it is not uncommon to observe linewidths that are 10--100 times larger than the ST prediction due to charge fluctuations \cite{Milonni1988}. In comparison, state-of-the-art lasers in atomic systems can reach, or even exceed, the ST limit to achieve linewidths close to 1 mHz \cite{Meiser2009, Kessler2012, Bohnet2012}. To harness the full potential of solid state masers and lasers, it is important to stabilize the emission frequency.

Injection locking is a commonly used method for narrowing the linewidth of a laser. With injection locking, laser emission is stabilized by the injection of an input tone that results in stimulated emission at the frequency of the injected tone. Frequency locking of oscillators has a rich history, extending back to Huygens in 1666, who observed that two initially unsynchronized clocks would eventually synchronize due to mechanical vibrations transmitted via a common beam \cite{Siegman1986}. Since then, frequency locking has been observed in systems ranging from fireflies \cite{Mirollo1990}, to spin transfer torque oscillators \cite{Kaka2005}, and matter waves of Bose gas \cite{Hofferberth2007}. For optical lasers, injection locking was first observed by Stover \textit{et al.}\ \cite{Stover1966}. It is now commonly used to improve the coherence of lasers by driving the input port with a narrow linewidth signal \cite{Schunemann1998}. Injection locking effects in a trapped ion phonon laser have been proposed for sensitive force sensing \cite{Herrmann2011}.  Other applications include amplification and frequency modulated signal detection in solid-state oscillators \cite{Kurokawa1973}. Laser combs with high power and coherence are generated by stabilizing a laser comb using the injection locking effect \cite{Fortier2006} and high resolution spectroscopy of cold atoms has been achieved by the self-injection of comb bunches \cite{DelHaye2014}.

In this paper we demonstrate injection locking of the recently discovered semiconductor DQD micromaser. The DQD micromaser is driven by single electron tunneling events between discrete zero-dimensional electronic states \cite{Liu2014}. A free-running emission linewidth of 34 kHz was measured, nearly 100 times larger than the ST prediction \cite{Schawlow1958}. Time-series analysis of the emitted signal indicates the maser output is fluctuating as a function of time. These fluctuations are believed to be due to charge noise, which electrostatically couples to the DQD energy levels and results in significant broadening of the emission peak.

Here we show that the emission linewidth can be narrowed by more than a factor of 10 using injection locking. For the case when the injected tone is detuned from the free running maser frequency by several linewidths, the maser emission frequency is ``pulled" by, and eventually locked to the input tone with increasing input power. The frequency range over which the maser can be injection locked $\Delta f_{\rm in}$ increases with injected signal power $P_{\rm in}$ following the power law relation $\Delta f_{\rm in}$  $\propto$  $\sqrt{P_{\rm in}}$ predicted by Adler \cite{Adler1946}. We also investigate the dynamics of the maser just outside of the injection locking regime, where the frequency pull is appreciable and leads to distortion sidebands in the emission spectrum. The emission powers and positions of the sidebands are in excellent agreement with theoretical predictions \cite{Siegman1986, Jahanpanah1996}. These measurements indicate that the DQD micromaser, which is driven by single electron tunneling events, follows predictions from conventional laser theory and can be considerably improved using injection locking effects.

\begin{figure}[t]
	\begin{center}
		\includegraphics[width=\columnwidth]{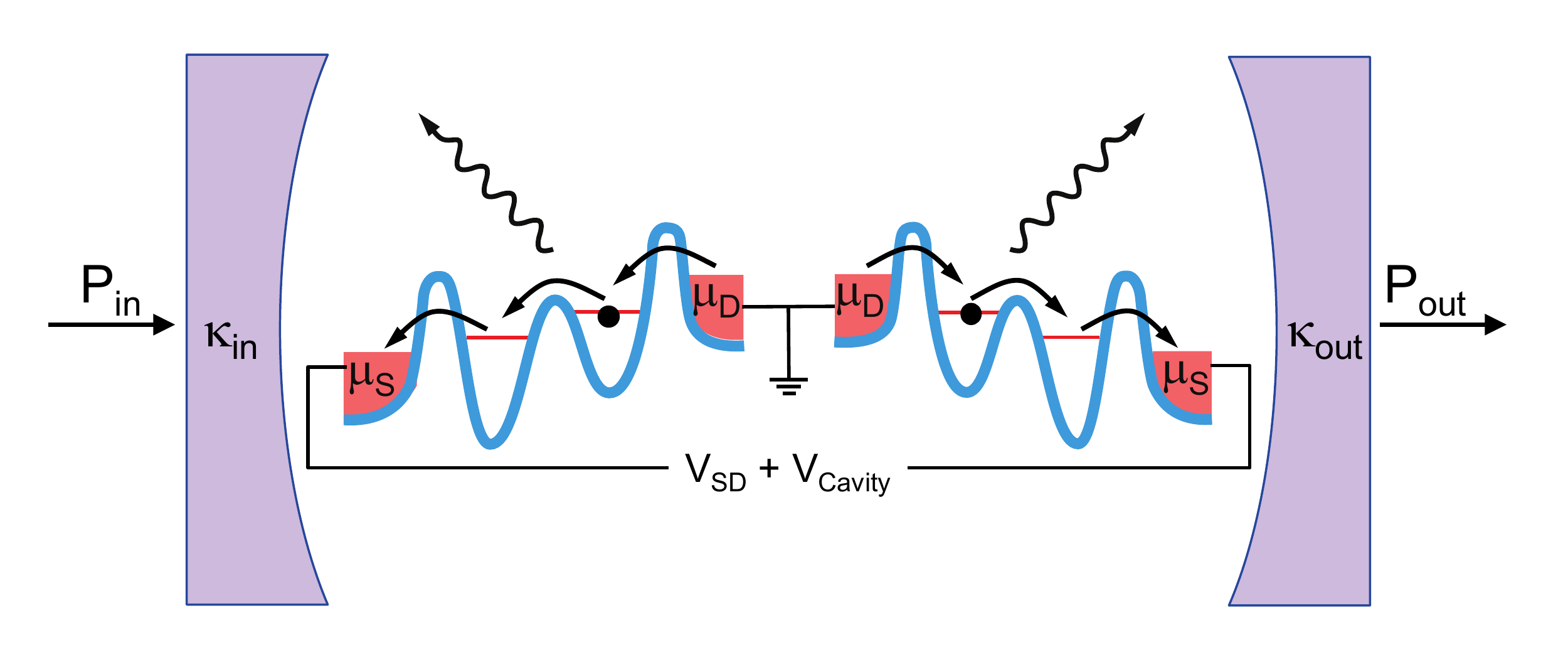}
		\caption{\label{sense1} Schematic of the DQD micromaser. Two DQDs are coupled to a high quality factor microwave cavity with input (output) coupling rates $\kappa_{\rm in}$ ($\kappa_{\rm out}$). A source-drain bias voltage $V_{\rm SD}$ = $(\mu_{\rm D} - \mu_{\rm S})/|e|$ results in single electron tunneling through the DQDs and leads to photon emission into the cavity mode.}
	\end{center}	
	%\vspace{-0.4cm}
\end{figure}

\section{Double Quantum Dot Micromaser}

We first briefly describe the main aspects of the DQD micromaser \cite{Liu2015}. The maser is fabricated in the circuit quantum electrodynamics architecture (cQED) and consists of a superconducting transmission line resonator, two semiconductor DQDs that serve as the gain medium, and a voltage bias that generates population inversion. The half-wavelength ($\lambda/2$) Nb coplanar waveguide resonator has a resonance frequency $f_c$ = 7880.6 MHz \cite{Wallraff2004, Frey2012, Petersson2012}. The cQED architecture has been used to achieve strong coupling between microwave frequency photons and a superconducting qubit \cite{Wallraff2004}. More recently, a variety of quantum dot devices (GaAs, carbon nanotubes, InAs nanowires, etched graphene) have been integrated with microwave cavities \cite{Frey2012, Toida2013, Petersson2012, Deng2013}.

The maser gain medium consists of two semiconductor DQDs, as illustrated in Fig.\ 1. Each DQD is fabricated by placing a single InAs nanowire across a predefined array of bottom gates \cite{Nadj-Perge2010}. Negative voltages are applied to the gates to selectively deplete the nanowire, forming a DQD \cite{Nadj-Perge2010, Wiel2002}. Electronic confinement results in a  discrete energy level spectrum \cite{Wiel2002, Fujisawa1998}. The energy levels can be tuned in-situ by adjusting the voltages applied to the bottom gates. In the nanowire DQDs investigated here, the electric dipole moment $d\sim 1000ea_0$, where $e$ is the electronic charge and $a_0$ is the Bohr radius \cite{Frey2012, Petersson2012}. We measure a charge-cavity interaction rate $g_c/2\pi \approx30$ MHz \cite{Petersson2012, Liu2014}. Quantum dots fabricated from other materials systems yield similar $g_c/2\pi$ = 10 -- 100 MHz \cite{Frey2012, Delbecq2011, Delbecq2013, Toida2013}.

Electron beam lithography is used to make source and drain contacts to the nanowires. A source-drain bias $V_{\rm SD}=2$ mV is applied to give a preferred direction for electron flow. Single electron tunneling is only allowed when the DQD energy levels are arranged such that an electron can tunnel downhill in energy [see Fig.\ 1], otherwise current flow is blocked due to Coulomb blockade \cite{Wiel2002}. Consider the left DQD shown in Fig.\ 1. Starting with an empty DQD, a single electron first tunnels from the drain to the right dot. This tunneling event is followed by an interdot charge transition from the right dot to the left dot, and subsequent tunneling of the electron from the left dot to the source. The source-drain bias effectively repumps the higher energy level in the DQD and generates conditions for population inversion. As shown in previous work, the interdot charge transition results in microwave frequency photoemission \cite{Liu2014,Liu2015}.

The DQD micromaser is in some ways similar to a quantum cascade laser (QCL). In a QCL, current flows through a precisely engineered quantum well structure and results in the cascaded emission of photons whose frequency is set by the quantum well layer thicknesses \cite{Faist1994}. In comparison, photons in the DQD micromaser are generated by single electron tunneling through electrically tunable DQD energy levels. While electrical control allows for \textit{in situ} tuning of the gain medium, it also means that the energy level separation will be susceptible to charge noise. To appreciate the magnitude of the noise, one can compare the measured root-mean-squared charge noise $\sigma_{\epsilon}/h = 10$ GHz with the much smaller $g_c/2\pi \approx30$ MHz and cavity linewidth $\kappa_{\rm tot}/2\pi \approx$  3 MHz \cite{Petersson2012, Liu2014,Frey2012, Toida2013, Petersson2012, Deng2013}. Charge noise will drive the DQDs out of resonance with the cavity, making it difficult to reach the strong-coupling regime \cite{Meschede1985, McKeever2003, Ates2009}. In terms of maser performance, charge fluctuations adversely impact the stability of the emission frequency and power \cite{Liu2015}.

Electron-phonon coupling is also an important factor in solid-state devices and has been extensively studied in semiconductor DQDs. Measurements of the inelastic current as a function of the DQD energy level detuning reveal oscillations that have been attributed to electron-phonon coupling \cite{Fujisawa1998, Weber2010}. These effects are especially pronounced when the dot size is comparable to the phonon wavelength \cite{Meunier2007}. The resulting orbital relaxation rate is on the order of 100 MHz \cite{Fujisawa1998, Petta2004}, again much larger than $g_c/2\pi$ and $\kappa_{\rm tot}/2\pi$. Recent studies of photoemission in cavity-coupled DQDs shows that only 1 photon is emitted into the cavity mode for every 1,000--10,000 electrons that tunnel through the DQD \cite{Liu2014, Stockklauser2015} and that the DQD maser gain profile can only be reproduced in theory when second order processes involving emission of photon and phonon are taken into account \cite{Gullans2015}. These previous studies highlight important differences between the DQD micromaser and its atomic beam counterparts \cite{Meschede1985}.

\section{Experimental Results}

We now present experimental data obtained on the semiconductor DQD micromaser. In Section III.A we briefly review recently published measurements of the DQD micromaser that examined the amplification of an input tone and measured the photon statistics of the maser in free-running mode (i.e.\ cavity emission in the absence of an input tone). These earlier works demonstrated that charge noise and phonons have a significant effect on maser operation \cite{Liu2015,Gullans2015}. In Section III.B we present new results showing that the maser emission can be injection locked by driving the input port of the cavity with a corresponding reduction in the emission linewidth. The injection locking range is measured as a function of input power and shown to be in good agreement with standard laser theory. Section III.C examines the frequency pull and distortion sidebands that appear outside of the locking range. Detailed analysis of the sidebands also yields excellent agreement with theoretical predictions.

\subsection{Free-Running Maser Characterization}

The maser is first characterized by driving the input port of the cavity at frequency $f_{\rm in}$ and power $P_{\rm in}$. Cavity power gain is defined as $G= C P_{\rm out}/P_{\rm in}$, where $P_{\rm out}$ is the power exiting the output port of the cavity. The normalization constant $C$ is defined such that the peak power gain $G_{\rm p}$ = 1 when both DQDs are configured in Coulomb blockade (off state). Figure 2 shows $G$ as a function of $f_{\rm in}$ with $P_{\rm in}$ = -120 dBm. The black curve is the cavity response in the off state \cite{Liu2015}. Fitting the gain to a Lorentzian we extract the cavity center frequency $f_c=7880.6$ MHz and linewidth $\kappa_{\rm tot}/2\pi=2.6$ MHz. Here $\kappa_{\rm tot}= \kappa_{\rm in} +\kappa_{\rm out} +\kappa_{\rm int}$. $\kappa_{\rm in}(\kappa_{\rm out})$ is the decay rate through the input(output) port and $\kappa_{\rm int}$ is the photon loss rate through other channels. The red curve shows $G$ as a function of $f_{\rm in}$ when current is flowing through both DQDs (on state). Here the cavity response is sharply peaked at $f_{\rm in}$ = 7880.25 MHz, yielding $G_{\rm p}\sim1000$ with a full-width-half-max (FWHM) $\Gamma$ = 0.07 MHz, suggestive of a transition to an above-threshold maser state.

Above-threshold maser action is confirmed by measuring the statistics of the output field as described in Ref.\ \cite{Liu2015}. These measurements are performed in free-running mode (with no input tone applied). The output signal is amplified and demodulated to yield the in-phase ($I$) and quadrature-phase ($Q$) components, which are sampled at a rate of 1 MHz. The results from 400,000 individual $(I,Q)$ measurements are shown in the two-dimensional histogram plotted in the inset of Fig.\ \ref{sense1}(b). The IQ histogram has a donut shape that is consistent with a stable oscillator, however the amplitude fluctuations are much larger than expected. Time-series analysis of the free-running emission signal suggests that large charge fluctuations are impacting the emission stability \cite{Liu2015}. It is therefore desirable to stabilize the output of the maser.

\begin{figure}[t]
	\begin{center}
		\includegraphics[width=\columnwidth]{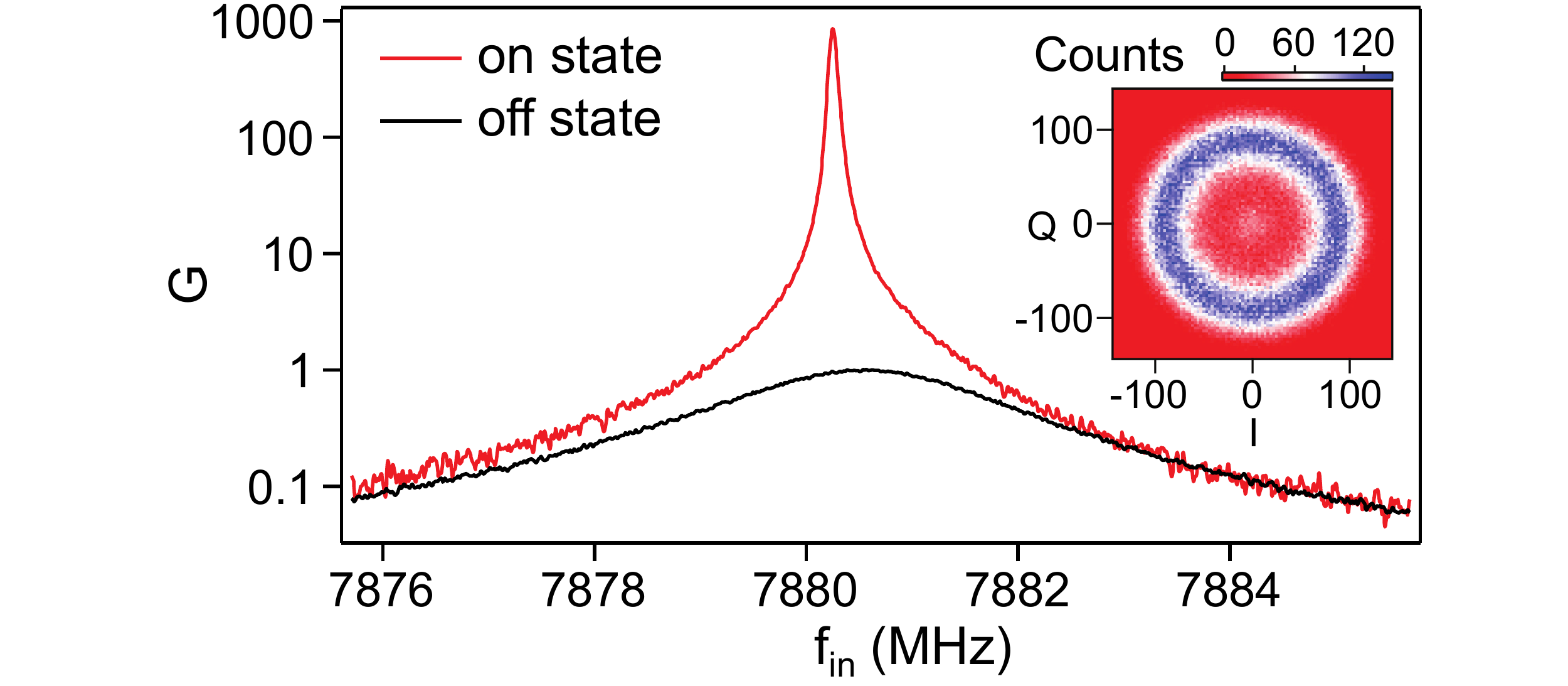}
		\caption{\label{sense1} Power gain $G$ = $CP_{\rm out}$/$P_{\rm in}$ plotted as a function of $f_{\rm in}$ with the DQDs configured in Coulomb blockade (off state) and with current flowing through the DQDs (on state). In the on state, the peak gain $G_{\rm p}\sim1000$ and the linewidth is dramatically narrowed, suggestive of a transition to a masing state. Inset: $IQ$ histogram of the output field measured in the on state (with no input tone applied to the cavity). The donut shape is indicative of above-threshold maser action \cite{Siegman1986, Liu2015}.}
	\end{center}	
	%\vspace{-0.4cm}
\end{figure}

\subsection{Injection Locking the Semiconductor DQD Micromaser}

\begin{figure*}[t]
	\begin{center}
		\includegraphics[width=2\columnwidth]{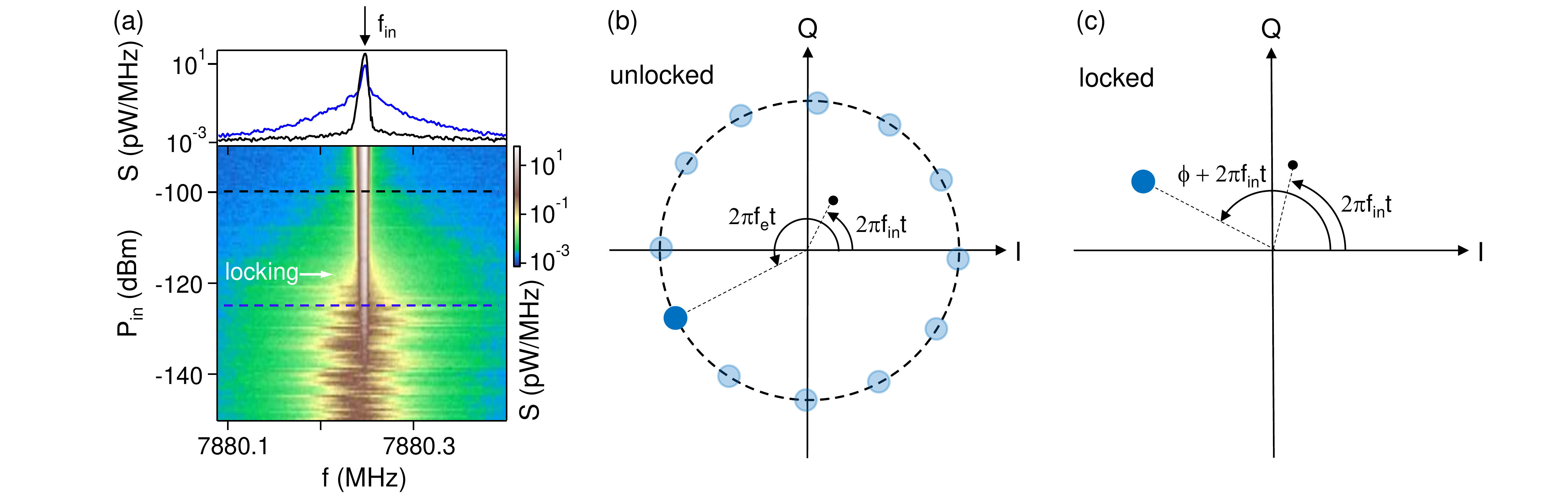}
		\caption{\label{sense2} (Color online) (a) Power spectrum of the emitted radiation $S(f)$ plotted as a function of $P_{\rm in}$. The cavity input frequency $f_{\rm in}$ = 7880.25 MHz is close to the free running maser emission frequency $f_{\rm e} = 7880.25 \pm 0.03$ MHz. Note the significant fluctuations in $f_{\rm e}$ for $P_{\rm in}$ $<$ -120 dBm. The maser linewidth narrows with increasing $P_{\rm in}$ due to injection locking. The upper panel shows $S(f)$ for $P_{\rm in}$ = -125 dBm (blue) and $P_{\rm in}$ = -100 dBm (black), indicated by the dashed lines in the main panel. (b) Phasor diagram of the maser output in the  unlocked configuration. Here the cavity field is a combination of the free running maser emission at frequency $f_{\rm e}$ and the cavity input tone at $f_{\rm in}$. In this configuration the phase of the maser is fluctuating relative to the input tone. (c) Schematic illustration of the cavity field in the injection locked state. To within a relative phase $\phi$, the maser emission is locked to the input tone.}
	\end{center}	
\end{figure*}

We now demonstrate injection locking of the maser by measuring the power spectral density of the emitted radiation $S(f)$ as a function of the power $P_{\rm in}$ of the input tone. The main panel of Fig.\ 3(a) shows $S(f)$ as a function of $P_{\rm in}$ with $f_{\rm in}$ = 7880.25 MHz set near the free running emission frequency $f_e$. Line cuts through the data are shown in the upper panel for $P_{\rm in} = -125$ dBm (blue curve) and $-100$ dBm (black curve). For negligible input powers ($P_{\rm in}<-140$ dBm) the power spectrum exhibits a broad peak near $f_{\rm e} = 7880.25$ MHz. For a given value of $P_{\rm in}$, the emission peak typically has a full-width-half-max (FWHM) $\Gamma$ = 34 kHz. For $P_{\rm in}$ $<$ -125 dBm, charge noise causes the emission peak to significantly wander in the frequency range $7880.25 \pm 0.03$ MHz. In this configuration the relative phases of the input tone and the maser emission are unlocked, as illustrated in Fig.\ 3(b). As $P_{\rm in}$ is increased, the photon number in the cavity at $f_{\rm in}$ increases, resulting in increased stimulated emission. With $P_{\rm in}$ $>$ -125 dBm, the broad tails of the emission peak are suppressed and the spectrum begins to narrow. The free running maser emission is eventually locked to the input tone around $P_{\rm in}=$ -115 dBm. Now the large fluctuations that were observed in the absence of an input tone are suppressed, and $\Gamma$ $<$ 3 kHz \cite{RBW}. The linewidth is reduced by more than a factor of 10 compared to the free-running case and indicates phase stabilization, as illustrated in Fig.\ 3(c).

Although our measurement of the linewidth is limited by technical effects \cite{RBW}, we can estimate the fundamental limit to the linewidth for this device.  Previous measurements on the output field indicated that the masing process intermittently shuts off due to large charge fluctuations that reduce the gain below threshold \cite{Liu2015}.   During these off periods, the maser emission will cease to be injection locked to the input tone and will lose phase coherence.  As a result, the linewidth will be limited by the inverse of the switching time $\tau_s$.  In Ref.\ \cite{Liu2015}, $\tau_s$ was observed to be roughly $500~\mu$s, which sets the fundamental linewidth limit due to charge noise as $\Gamma \sim 1/\tau_s \approx 2~$kHz.  The linewidth prediction is comparable to our measurement resolution and a factor of 10 smaller than the linewidth of the free-running maser, but still larger than the ST limit by the same factor.  Further reductions of the linewidth will most likely require reducing charge noise in these devices.

Comparable effects are observed when $f_{\rm in}$ = 7880.60 MHz, more than 10 line-widths detuned from $f_{\rm e}$ [Fig.\ 4]. With $P_{\rm in}<-140$ dBm only the free running emission peak is visible in $S(f)$. As $P_{\rm in}$ is further increased the injection tone becomes visible and the power spectrum is simply a sum of the free running maser emission and the cavity input tone. When $P_{\rm in}$ $\gtrsim$ -125 dBm distortion sidebands appear and the free running emission peak is pulled towards the input tone. The maser abruptly locks to $f_{\rm in}$ when $P_{\rm in}=$ -102 dBm, but the emission is still somewhat broad. The linewidth continues to narrow until $P_{\rm in}=$ -98 dBm, beyond which point the measured linewidth is limited by experimental factors \cite{RBW}. The upper panel of Fig.\ 4 shows line cuts through the data, acquired at $P_{\rm in}=$ -115 dBm (red curve) and $P_{\rm in}=$ -100 dBm (black curve). The sidebands that are visible in $S(f)$ (marked $n$ = -2, 0, 1, and 2) are quantitatively analyzed in Section III.C \cite{Siegman1986, Jahanpanah1996}.

\begin{figure}[t]
	\begin{center}
		\includegraphics[width=\columnwidth]{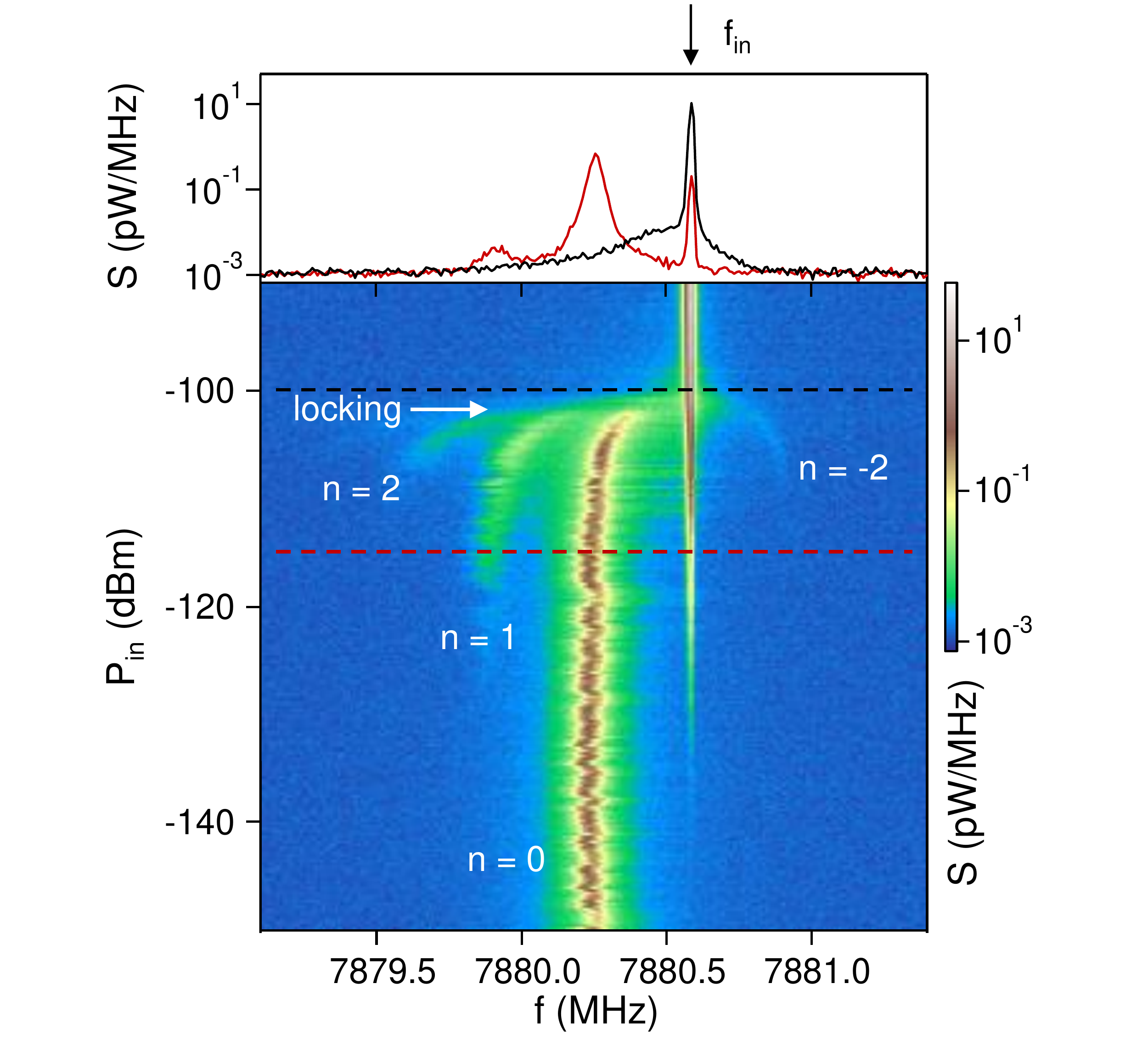}
		\caption{\label{sense3} (Color online) $S(f)$ plotted as a function of $P_{\rm in}$ with $f_{\rm in}$ = 7880.6 MHz far detuned from $f_{\rm e}$ (note the change in the x-axis scale relative to Fig.\ 3).  The maser is injection locked when $P_{\rm in}$ $>$ -102 dBm. Distortion sidebands are clearly visible in the emission spectrum. Upper panel: $S(f)$ for $P_{\rm in}$ = -115 dBm (red) and $P_{\rm in}$ = -100 dBm (black), indicated by the dashed lines in the main panel.}
	\end{center}
	\vspace{-0.4cm}
\end{figure}

We next measure the frequency range over which the maser is injection locked. The upper inset of Fig.\ 5 shows a color-scale plot of $S(f)$ as a function of $f_{\rm in}$ measured with $P_{\rm in}$ = -110 dBm. The input signal is visible in $S(f)$ and marked with an arrow for clarity. As seen in the data, $f_{\rm in}$ has little effect on the maser emission when it is far-detuned from $f_{\rm e}$. As $f_{\rm in}$ is increased and brought closer to $f_{\rm e}$, frequency pulling is visible and emission sidebands appear. The maser then abruptly locks to $f_{\rm in}$, and remains locked to $f_{\rm in}$ over a frequency range $\Delta f_{\rm in}$ = 0.27 MHz. The lower inset of Fig.\ 5 shows $S(f)$ as a function of $f_{\rm in}$ with $P_{\rm in}$ = -100 dBm. Here the maser is injection locked over a larger range $\Delta f_{\rm in}$ = 0.85 MHz. Similar to the upper inset, frequency pulling and sidebands are observed outside of the injection locking range. By repeating these measurements at different $P_{\rm in}$, we obtain the data shown in the main panel of Fig.\ 5, where $\Delta f_{\rm in}$ is plotted as a function of $P_{\rm in}$. The blue line in Fig.\ 5 is a fit to the power law relation $\Delta f_{\rm in}$ = $A_{\rm M}$ $\sqrt{P_{\rm in}}$, with the measured prefactor $A_{\rm M}$ = $ (2.7 \pm 1.0) \times 10 ^6\; {\rm MHz/\sqrt{W}}$, where the error bar is due to 3 dB of uncertainty in the transmission line losses.

The measured power law relation can be compared with predictions from Adler's theory, which considers the maser dynamics in the rotating frame of the input tone by assuming that the input power is small compared to the free emission power \cite{Adler1946}.  We express the cavity output field as
\begin{equation}
\alpha(t)=I(t) +i Q(t)=\sqrt{P_e} e^{2 \pi i f_{\rm in} t+ i \phi(t)},
\end{equation}
 where $P_e$ is the emitted power (assumed to be constant) and  ${\phi=\phi_{\rm e}- \phi_{\rm in}}$ is the relative phase of the input field $\phi_{\rm in}$ and the emitted field $\phi_{\rm e}$.  The relative phase follows the Adler equation:
\begin{equation}
\frac{d\phi}{dt}+2\pi(f_{\rm in}-f_{\rm e}) =-2\pi\frac{\Delta f_{\rm in}}{2}\sin(\phi).
\label{Eq: Adler equation}
\end{equation}

\noindent In the injection locking range $|f_{\rm in}-f_{\rm e}|<\Delta f_{\rm in}/2$, Eq.\ (\ref{Eq: Adler equation}) has a static solution $\phi=\arcsin \left[2(f_{\rm e}-f_{\rm in})/\Delta f_{\rm in}\right]$. The emission phase is then ``locked" to the input tone with $\phi\in(-\pi/2 , \pi/2)$, which corresponds to the case illustrated in Fig.\ 3(c).

\begin{figure}
	\begin{center}
		\includegraphics[width=\columnwidth]{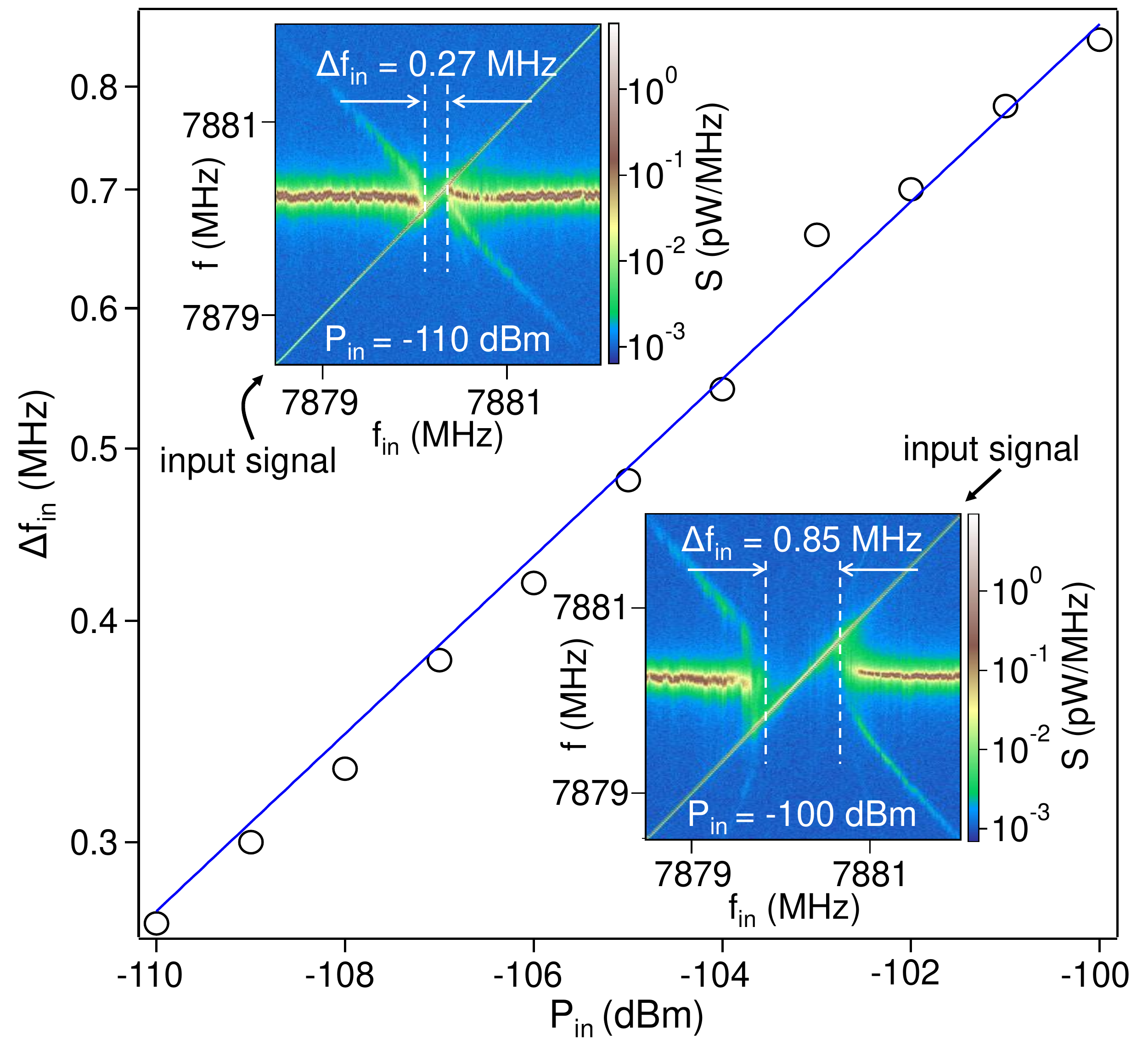}
		\caption{\label{sense4} (Color online)
			The frequency range over which the maser is injection locked, $\Delta f_{\rm in}$, increases with $P_{\rm in}$. The blue line is a fit to the power law $\Delta f_{\rm in} \propto \sqrt{P_{\rm in}}$ prediction of the Adler equation. Insets: $S(f)$ measured as a function of $f_{\rm in}$ with $P_{\rm in}$ = -110 dBm (upper left) and $P_{\rm in}$ = -100 dBm (lower right).}
	\end{center}
	\vspace{-0.4cm}
\end{figure}

Adler's analysis shows that $\Delta f_{\rm in}$ is proportional to the amplitude of the input signal such that
\begin{equation}
	\Delta f_{\rm in} = C_\kappa \frac{\kappa_{\rm tot}}{2\pi} \sqrt{P_{\rm in}/P_{\rm e}} \equiv A_{\rm T} \sqrt{P_{\rm in}}.
	\label{Eq: locking range}
\end{equation}
The cavity prefactor $C_\kappa =  2\sqrt{\kappa_{\rm in}\kappa_{\rm out}}/\kappa_{\rm tot}$ accounts for internal cavity losses and is obtained using cavity input-output theory \cite{Siegman1986}. Our microwave cavity is designed with $\kappa_{\rm in}/2\pi = \kappa_{\rm out}/2\pi=0.39$ MHz, and $\kappa_{\rm tot}/2\pi =2.6$ MHz is directly extracted from the data in Fig.\ 2. These quantities yield $C_{\kappa} = 0.3$. The average emitted maser output power $P_{\rm e}$ $\approx$ $(2.5 \pm 1.9) \times 10 ^{-2}$ pW. Using these quantities we find
\begin{equation*}
	A_{\rm T} = \frac{C_\kappa}{\sqrt{P_{\rm e}}} \frac{\kappa_{\rm tot}}{2\pi} = \left(4.9 \pm 1.7 \right) \times 10 ^6\; {\rm MHz/\sqrt{W}}.
\end{equation*}
\noindent We therefore find reasonable agreement between the data and the predictions from Adler's theory, considering the uncertainties in the transmission line losses.

\subsection{Behavior Outside of the Injection Locking Range: Frequency Pull and Distortion Sidebands}
\label{Sec: sidebands}

\begin{figure*}
	\begin{center}
		\includegraphics[width=2\columnwidth]{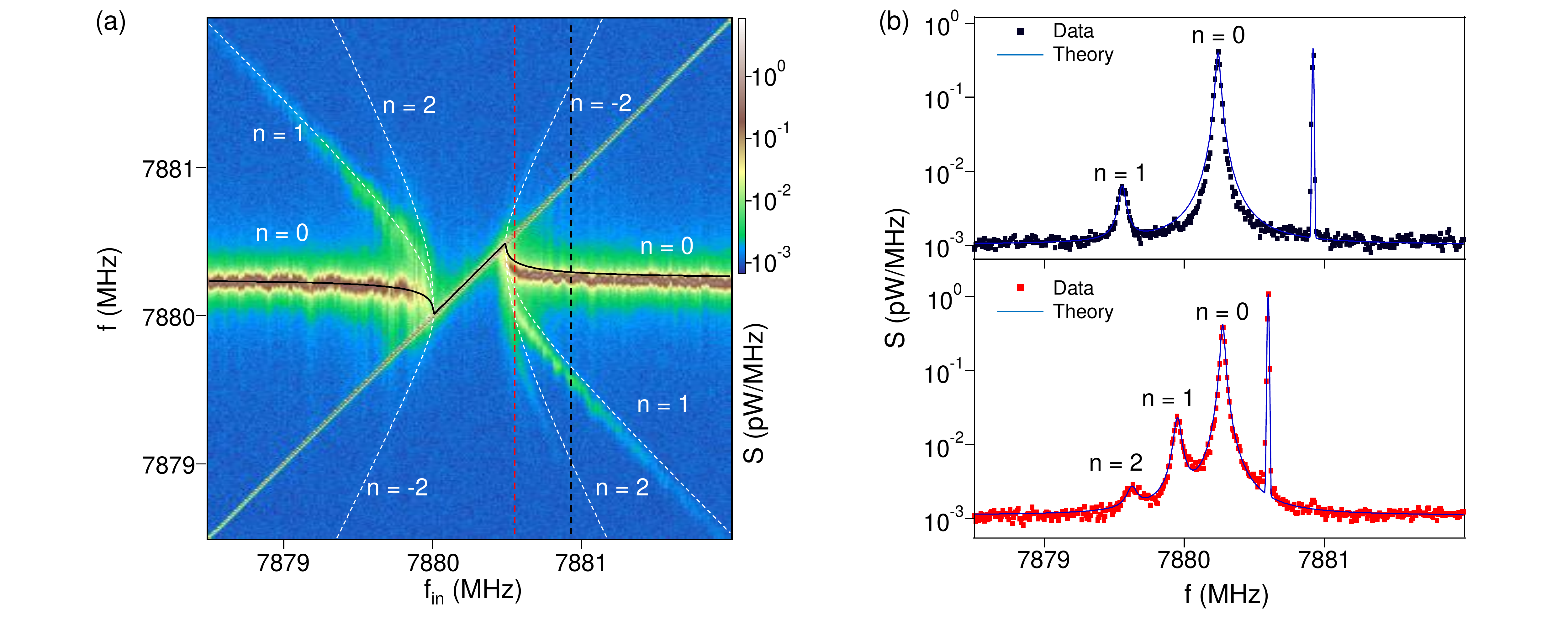}
		\caption{\label{sense5} (Color online)
			(a) $S(f)$ measured as a function of $f_{\rm in}$ with $P_{\rm in}$ = -105 dBm. The black line overlaid on the data is the “pulled” frequency predicted by Adler’s injection locking theory. The white dashed lines are the predicted sideband locations. (b) $S(f)$ for $f_{\rm in}$  = 7880.92 MHz in (upper panel) and $f_{\rm in}$  = 7880.60 MHz (lower panel), indicated by the dashed lines in (a). Solid lines are fits to $S(f)$.}
	\end{center}
	\vspace{-0.4cm}
\end{figure*}

We now examine the behavior of the maser outside of the injection locking range, where the frequency pull is appreciable and distortion sidebands are visible. Figure\ \ref{sense5}(a) shows $S(f)$ as a function of $f_{\rm in}$ with $P_{\rm in} = -105$ dBm. Injection locking is observed over a frequency range $\Delta f_{\rm in}$ = 0.48 MHz. Focusing on the region with $f_{\rm in} > 7880.5$ MHz, we observe one sideband for $f > f_{\rm e}$ and two sidebands for $f < f_{\rm e}$. For clarity, the emission peaks are labeled with the index $n$: $n=0$ corresponds to the frequency pulled maser emission peak, $n=-1$ corresponds to the input tone, and the other peaks are distortion sidebands. Figure\ \ref{sense5}(b) shows line cuts through the data at $f_{\rm in}$ = 7880.92 MHz (upper panel) and $f_{\rm in}$ = 7880.60 MHz (lower panel). When $f_{\rm in}$ = 7880.92 MHz, the pulled emission peak $\bar{f}_{\rm e}$ ($n$ = 0) is detuned from $f_{\rm in}$ by the beat frequency $f_{\rm b} = \bar{f}_{\rm e}- f_{\rm in}$.
For this set of parameters we measure $f_{\rm b} = -0.68$ MHz. The $n$ = 1 sideband is detuned from the $n$ = 0 peak by $f_{\rm b}$. When $f_{\rm in}$ = 7880.60 MHz, the $n$ = 2 sideband is also visible. To allow for a quantitative comparison with theory, we analyze the spectra in Fig.\ 6(b) by fitting the sideband emission peaks to a Lorentzian lineshape and the input tone to a Gaussian with a width of 10 kHz \cite{RBW}. The sideband powers $P_{\rm n}$ are listed in Table\ \ref{Table peak}.

To compare the data with theory, we seek a general solution for $\phi(t)$. In the limit of small $P_{\rm in}$, $\Delta f_{\rm in} \approx 0$ and $\phi(t) \approx 2\pi(f_{\rm e}-f_{\rm in})t$. In this case the cavity field can simply be considered as a sum of the free emission signal and the cavity input tone, as shown in Fig.\ 3(b). Outside of this limit, we solve the Adler equation analytically to find the cavity field
\begin{equation}
\alpha = \sqrt{P_{\rm e}}e^{2\pi i(f_{\rm in} + f_{\rm b})t}\left(\sum_{n = -\infty}^{\infty} a_n e^{2\pi i n f_{\rm b}t}\right).
\label{Eq: total caivty field}
\end{equation}
The expansion coefficients $a_n$ have been calculated by Armand and are given later \cite{Armand1969}. The beat frequency is found self-consistently from this solution
\begin{equation}
f_{\rm b} = (f_{\rm e}-f_{\rm in})\sqrt{1-\left(\frac{\Delta f_{\rm in}/2}{f_{\rm e}-f_{\rm in}}\right)^2}.
\label{Eq: injection pull}
\end{equation}
Given that $f_{\rm e}$ wanders in the frequency range $7880.25 \pm 0.03$ MHz, Eq.\ \ref{Eq: injection pull} predicts $f_{\rm b} = -0.63 \pm 0.03$ MHz at $f_{\rm in} = 7880.92$ MHz and, $f_{\rm b} = -0.25 \pm 0.03$ MHz at $f_{\rm in} = 7880.60$ MHz. These values are in general agreement with the measured $f_{\rm b}$ listed in Table\ \ref{Table peak}. The small discrepancy may be due to charge-noise-induced drift in $f_{\rm e}$.

Predicted sideband positions can be obtained by evaluating Eq.\ \ref{Eq: total caivty field} in several different regimes. For the far detuned case $\left|  f_{\rm e}-f_{\rm in} \right| \gg \Delta f_{\rm in}$, higher order harmonics are negligible and $a_0\approx 1$.  Equation (\ref{Eq: total caivty field}) then simplifies to  $\alpha = \sqrt{P_e}e^{2\pi i(f_{\rm in} +  f_{\rm b})t}$, which represents the pulled emission peak at frequency $\bar{f}_e = f_{\rm in} + f_b$. First order expansion of Eq.\ \ref{Eq: total caivty field} in $\Delta f_{\rm in}$ yields $a_{\pm 1} \approx i\Delta f_{\rm in}/4(f_{\rm e}-f_{\rm in})$ \cite{Stover1966a}. When the detuning $\left| f_{\rm e}-f_{\rm in} \right|$ approaches $\Delta f_{\rm in}/2$, higher order terms in Eq.\ \ref{Eq: total caivty field} give rise to non-negligible expansion coefficients $a_n$, which results in higher order sideband peaks at frequencies $f_n = \bar{f}_e \pm n f_b$ ($n$ = $\pm1$, $\pm2$, ...). The predicted $\bar{f}_e$ is plotted as black solid line in Fig.\ \ref{sense5}(a) and the predicted $n$ $\neq$ 0 sidebands are plotted as white dashed lines. Both the ``pulled" emission peak and the location of the distortion sidebands are in good agreement with Adler's theory.

The sideband powers $P_{\rm n}$ can be compared with calculations from Armand \cite{Armand1969}, who found: %predict:
\[a_{-1} = \frac{ f_{\rm e}-f_{\rm in}-f_{\rm b}+i(\Delta f_{\rm in}/2)}{ f_{\rm e}-f_{\rm in}+f_{\rm b}-i(\Delta f_{\rm in}/2)}.\]
Since the $n$ = -1 sideband overlaps with the input tone, it cannot be resolved experimentally. $a_n$ = 0 for $n$ $\leq$ -2,
\[a_0 =\frac{ {4 (f_{\rm e}-f_{\rm in})f_{\rm b}} }{\left[ f_{\rm e}-f_{\rm in}+f_{\rm b}-i(\Delta f_{\rm in}/2)\right] ^{2}},
\]
and for $n>0$
\begin{equation}
{a_n}={a_0}\left[ \frac{\left(-f_{\rm e}+f_{\rm in}+f_{\rm b}\right)+i\left(\Delta f_{\rm in}/2\right)}{\left( f_{\rm e}-f_{\rm in}+f_{\rm b}\right)-i\left(\Delta f_{\rm in}/2\right)}\right]^n.
\label{Eq: sideband amplitude ratio}
\end{equation}
The predictions imply that the $n$ $>$ 0 sidebands are favored, an asymmetry that is consistent with the data in Fig.\ \ref{sense5}(b), as well as other laser systems \cite{Stover1966a, St-Jean2014}.  We can understand this at a qualitative level by noting that, outside the injection locking region, $\dot{\phi}$ has a fixed sign equal to the sign of $(f_{\rm e}-f_{\rm in})$.  As a result, the maser signal is always trailing or leading the input signal in phase.  Since $f_{\rm in} - \dot{\phi}$ is the approximate frequency of the oscillator, this implies that the formation of sidebands will always be favored at higher(lower) frequencies with respect to $f_{\rm in}$ when $f_{\rm e}$ is greater(smaller) than $f_{\rm in}$.

\begin{table}
	\begin{center}
		%\vspace{0.5 cm}
		\caption{Distortion Sideband Parameters}
		\begin{ruledtabular}
			\begin{tabular}{l  l  l }
				$f_{\rm in}$ (MHz) & 7880.92 &  7880.60 \\\hline
				$f_{\rm b}$ (MHz) & -0.68 &  -0.33 \\
				$P_0$  (pW) & $2.1\times10^{-2}$ & $2.0\times10^{-2}$ \\
				$P_1$  (pW) & $5.4\times10^{-4}$ & $1.8\times10^{-3}$ \\
				$P_2$  (pW) & NA & $1.59\times10^{-4}$ \\
                $P_1/P_0$ & $2.7\times10^{-2}$ & $9.1\times10^{-2}$ \\
                $P_2/P_0$ & NA & $7.9\times10^{-3}$ \\
				$|a_1/a_0|^2$ & $2.9\times10^{-2}$ & $1.0\times10^{-1}$ \\
				$|a_2/a_0|^2$ & NA & $1.1\times10^{-2}$
				\label{Table peak}
			\end{tabular}
		\end{ruledtabular}
	\end{center}
\end{table}

We can now compare the measured sideband powers with the theoretical predictions. For the data shown in the upper panel of Fig.\ \ref{sense5}(b), we find $P_1/P_0 = 2.7\times10^{-2}$, which is very close to the value predicted by Eq.\ \ref{Eq: sideband amplitude ratio} $|a_1/a_0|^2 = 2.9\times10^{-2}$ . The theoretical value is calculated taking the measured beat frequency $f_{\rm b}$ = -0.68 MHz, the measured $\Delta f_{\rm in}$ = 0.48 MHz obtained with $P_{\rm in} = -105$ dBm, and $f_{\rm e}-f_{\rm in}=$ -0.72 MHz determined from Eq.\ \ref{Eq: injection pull}. Similarly, for the lower panel of  Fig.\ \ref{sense5}(b) Adler's theory predicts ratios $|a_1/a_0|^2$ = 1.0$\times10^{-1}$ and $|a_2/a_0|^2$ = $1.1\times10^{-2}$, which are also in good agreement with the experimental results listed in Table\ \ref{Table peak}.

\section{Conclusion and Outlook}

%summarize the results of the experiment.
In conclusion, the emission linewidth of the semiconductor DQD micromaser can be narrowed by more than a factor of 10 using injection locking. Measurements of the injection locking range as a function of input power very closely follow predictions from Adler's theory \cite{Adler1946}. We also examined the frequency pull and emission sidebands outside of the injection locking regime. Our data show that this exotic maser, which is driven by single electron tunneling events, is well-described by predictions from conventional laser theory. Future areas of work include the development of a quantitative theory to explain how charge noise impacts the emission peak location and linewidth, steps to improve materials to reduce charge noise, and investigation of the micromaser in the single emitter limit (with one semiconductor DQD in the cavity).

\begin{acknowledgments}
Research at Princeton was funded in part by the Packard Foundation, the National Science Foundation (Grants No.\ DMR-1409556 and DMR-1420541), and the Gordon and Betty Moore Foundation’s EPiQS Initiative through Grant GBMF4535.
\end{acknowledgments}

%\bibliographystyle{apsrev}
%\bibliography{lasing_nourl}

\begin{thebibliography}{47}
\bibitem [{RBW()}]{RBW}%
  \BibitemOpen
  \href@noop {} {}\bibinfo {note} {The frequency resolution is limited by the
  resolution bandwidth of the Agilent spectrum analyzer E4405B. For Fig.\ 3 the
  RBW is set to 3 kHz and 10 kHz for Fig.\ 4-6.}\BibitemShut {Stop}%
\end{thebibliography}

%

\end{document}